\documentclass{elsart}
 
\begin{document}
 
  
\begin{frontmatter}

\title{The wavelength of neutrino and neutral kaon oscillations}
   
\author{H. Burkhardt\thanksref{notts},}
\author{J. Lowe and G.J. Stephenson Jr.} 
\address{Physics and Astronomy Department, University of 
New Mexico, Albuquerque, \\NM 87131, USA}
  
\author{T. Goldman}
\address{Theoretical Division, Los Alamos National Laboratory, 
Los Alamos, \\NM 87545, USA}
 
\thanks[notts]{Also
at Shell Centre for Mathematical Education, University, Nottingham
NG7 2RD, England}

\begin{abstract}
Neutral kaons, and probably also neutrinos, exhibit oscillations 
between flavor eigenstates, as a result of being produced in a 
superposition of mass eigenstates.  Several recent papers 
have addressed the question of the energies and momenta of the 
components of these states, and their effect on the coherence of the
states and on the wavelength of the oscillations. We point
out that the mass eigenstates need have neither equal momentum nor
equal energy, but can 
nevertheless be coherent, and that a correct treatment of the 
kinematics recovers the usual result for the wavelength of the 
flavor oscillations.
\end{abstract}
\end{frontmatter}
 
\section{Introduction}
 
When neutral particles are produced in a flavor eigenstate that 
is not also a mass eigenstate, the resultant system is, in 
general, a superposition of mass eigenstates. If so, the system may 
oscillate between the different flavor eigenstates. This 
situation has been
familiar for many years for the case of neutral kaons; if these
are produced in one of the strangeness eigenstates, $K^0~(S=1)$
or $\bar{K}^0~(S=-1)$, the system is a superposition of the mass
eigenstates, $K_L$ and $K_S$, and oscillates between the two 
strangeness eigenstates. Recently, strong evidence has been found
that neutrinos show a similar behaviour 
\cite{lsnd,superk,sno,kam}.
  
The standard quantum-mechanical treatment of kaon oscillations
\cite{bj} has been known for many years and results in an
expression relating the wavelength of the strangeness oscillations
to the mass difference between the mass eigenstates,
$\delta m=m_L-m_S$. In the last decade, several papers have appeared
which question this treatment, sometimes resulting in a different
relation between the wavelength and $\delta m$. Srivastava 
et al. \cite{sriv} derived a relation that 
differs by at least a factor of 2 from the standard result. The origin 
of this factor was studied by Lowe et al. \cite{jlowe} 
and by Burkhardt et al. \cite{hb} who found an error
in Refs. \cite{sriv}, and demonstrated that the standard 
result is recovered when this error is corrected.
 
Other treatments 
\cite{lip1,kayser1,kayser2,stod,lip2,okun1,dolgov,okun2} 
have studied some consequences of differing assumptions about the system 
and, in particular, the energies and momenta of the mass eigenstates. 
Since the $K_L$ and $K_S$ states that
make up the oscillating $K^0-\bar{K}^0$ system have different
masses, they cannot have {\em both} the same momentum {\em and}
the same energy. For example, Lipkin and collaborators 
\cite{lip1,lip2} have studied the consequences of assuming either 
equal momentum or equal energy for the $K_L$ and $K_S$. In several
of the above papers these two kinematic assumptions are examined,
and some papers predict a 
wavelength for the oscillations which differs by a factor
of exactly 2 from the standard treatment. A recent paper by Okun 
et al. \cite{okun2} gives a concise summary of the situation.
 
However, we are not free to choose the energy and momentum of the
mass eigenstates. Usually, the neutral particles are produced
either in a reaction (e.g. $\pi^- p\rightarrow \Lambda K^0$) for
the case of kaons, or a decay (e.g. $\pi\rightarrow\mu\nu_{\mu}$)
for neutrinos. In either case, the mass eigenstates have neither
the same energy nor the same momentum for a given center-of-mass
energy (mass in the rest frame of the source). The energies and 
momenta are determined by the kinematics. This was pointed out 
initially
by Boehm and Vogel \cite{bv}, Goldman \cite{tg}, Srivastava 
et al. \cite{sriv}, Dolgov \cite{dolgov} and also by 
the present authors \cite{jlowe,hb}, who showed that when this is 
taken into account correctly, a consistent treatment of the kinematics 
follows and the standard result for the wavelength of oscillations 
is recovered.
 
However, Lipkin et al. \cite{lip1,lip2} and Stodolsky 
\cite{stod} claim that the two mass eigenstates must have the same 
energy. In particular, Lipkin \cite{lip2} states that mass eigenstates
with different energy cannot interfere to produce the oscillations. 
In sect. \ref{sect:coh} we show that this argument is incorrect and 
that the $K_L$ and $K_S$ states can indeed interfere to give 
strangeness oscillations. In sect. \ref{sect:kin}, 
we show that the mass eigenstates with kinematically correct
energies and momenta produce oscillations with the same
wavelength as in the standard treatment, without any additional 
factors.
   
\section{Coherence of the mass eigenstates} \label{sect:coh}
 
In this section, we examine the coherence of two interfering wave
functions. Suppose the wave functions are plane waves, $\psi_1$
and $\psi_2$. These might be, for example, two parts of the wave
function in a 2-slit optical experiment, or an electron diffraction 
experiment. If so, they will have the same energy and momentum,
but for generality, we keep both energies and momenta distinct
for now.

The wave function at the point of interference, $(x,t)$, is
 
\begin{eqnarray}
\psi=\psi_1+\psi_2=\sqrt{\frac{1}{2}}\bigg[
{\rm exp}\{{\rm i}(p_1x-E_1t)\}+
{\rm exp}\{{\rm i}(p_2x-E_2t+\phi)\}\bigg]
\label{eq:eq1}
\end{eqnarray}
  
\noindent where $\phi$ is some phase angle introduced by the 
geometry (for example, the path difference between the two slits 
to the interference point or some difference induced at the 
presumed common source of the two components). We assume only 
that $\phi$ is fixed and does not vary, e.g., randomly
for different times due, for example, to fluctuations in the
background medium (e.g., due to index of refraction 
fluctuations). It is in this sense that we may refer to the two
waves as ``coherent" \cite{zurek}. The probability density is 
 
\begin{eqnarray}
\mid\psi\mid^2=1+
{\rm cos}[(p_1-p_2)x-(E_1-E_2)t+\phi]
\label{eq:eq2}
\end{eqnarray}
 
\noindent In general, the second term oscillates in time 
with a time period characteristic of the energy scale of the
system. For the optical case, this is $\sim 10^{-15}$s, and in
particle-physics experiments, the characteristic time is much
shorter. In either case, this is well below the time resolution 
of any normal detector, so the second term averages to zero in 
the measurement, and $\psi_1$ and $\psi_2$ are therefore 
incoherent in the sense of Lipkin et al. \cite{lip1,lip2}. 
There may be other reasons for the cross term in 
Eq. (\ref{eq:eq2}) to vanish
(e.g. different spin wave functions or different internal states
of particles associated with $\psi_1$ and $\psi_2$), but in the
absence of any such reason, the rapid time dependence of the
energy term is the only reason why $\psi_1$ and $\psi_2$
are orthogonal when $E_1$ and $E_2$ are different. It is this
energy term that seems to be the basis for statement often
made (e.g. in \cite{lip2}) that the kaon mass eigenstates are
incoherent, and will not interfere, unless they have the same 
energy.
 
However, we must examine Eq. (\ref{eq:eq2}) in the particular case 
where $\psi_1$ and $\psi_2$ are $K_L$ and $K_S$ states. Here,
$E_1-E_2$ is of order $\delta m$, which is 
$\sim 3\times 10^{-6}$eV. The associated time scale is
$\sim 2\times 10^{-10}$s, which is readily measurable. 
Thus, even in this plane-wave case (that is, in the absence of
packeting), the response of a detector to $\mid\psi\mid^2$ at a
fixed point $x$ will oscillate measurably in time as given by
Eq. (2). Furthermore, in actual experiments, the kaons do not appear 
in continuous plane waves, but as wave packets. Since 
kaon velocities in experiments are usually an appreciable
fraction of $c$, the distance scale is many cm, again readily 
measurable. Of course, if the measurement averages
over time or distance scales large compared with these values,
then the cross term vanishes and the states become incoherent.
 
Thus, although the fact that the kinematics of the kaon (or
neutrino) production process preclude equal energy for the
two interfering states, the states may nevertheless be coherent, 
in the sense that we refer to above,
and interference may be observable in certain situations.
 
\section{Kinematics} \label{sect:kin}
 
Here, we calculate the kinematics for a specific case. For
definiteness, we choose the $K^0-\bar{K}^0$ example rather
than neutrinos because

\noindent (i) There are just two states rather than three or more,
 
\noindent (ii) Accurate numerical values are known for the masses 
and the mass difference, 
 
\noindent (iii) For neutrinos, it has been argued that a full 
treatment requires the inclusion of the detector in the system 
\cite{lip2}. However, for kaons, the $\Delta S=\Delta Q$ rule 
implies that the $K^0$ and $\bar{K}^0$ components can be 
identified from the kaon decay, without the need for a specific 
detector, thus simplifying the problem.

Neutrinos or kaons are produced either by a reaction such as
 
$$\pi p\rightarrow\Lambda K^0$$
 
\noindent or by a decay, for example
 
$$\pi\rightarrow\mu\nu_{\mu}.$$
 
\noindent For the first of these, the center-of-mass energies and 
momenta of the mass eigenstates are given by
 
\begin{eqnarray}
p_i^2=\frac{(s-m_i^2-m_{\Lambda}^2)^2-4m_i^2m_{\Lambda}^2}{4s},~~~~
E_i=\frac{s+m_i^2-m_{\Lambda}^2}{2\sqrt{s}}
\label{eq:eq3}
\end{eqnarray}
 

\noindent where $i=S$ or $L$. Similar expressions hold for the 
neutrinos from pion decay. The quantity 
$\sqrt{s}$ is the total center-of-mass energy for a reaction or 
the mass of the decaying particle for a decay. Although there
may be a spread in the value of $\sqrt{s}$ for the overall
system wave packet, our analysis proceeds component by
component, i.e. at a precise value of $\sqrt{s}$
(within the constraints of the uncertainty principle). Thus 
$p_S\neq p_L$ and $E_S\neq E_L$. In the following, to make the 
equations more readable, we ignore CP violation and we omit the 
widths of the kaon states. 
 
Since the above reaction produces a pure $K^0$ state,
the wave function at the reaction point, where $x=t=0$, is
 
\begin{eqnarray}
\mid K^0\rangle=\sqrt{\frac{1}{2}}(\mid K_L\rangle+
\mid K_S\rangle).
\label{eq:eq4}
\end{eqnarray}
 
\noindent This state develops in time as 
 
\begin{eqnarray}
\psi(x,t)=\sqrt{\frac{1}{2}}\{{\rm exp}[{\rm i}(p_Lx-E_Lt)]
\mid K_L\rangle+{\rm exp}[{\rm i}(p_Sx-E_St)]\mid K_S\rangle\}.
\label{eq:eq5}
\end{eqnarray}
 
\noindent Since Eqs. (\ref{eq:eq3}) 
give $p_i$ 
and $E_i$ in the center-of-mass frame, the $x$ and $t$ in 
Eq. (\ref{eq:eq5}) should also be in this frame. However, 
Eq. (\ref{eq:eq5}) and all following equations
involve only invariants, so may be reinterpreted in the lab
frame. Note that the two components in Eq. 
(\ref{eq:eq5}) are coherent in 
the sense that we defined above: no fluctuation of the relative 
phase occurs at the source. Under the assumption of propagation in
a vacuum, there is also no medium to induce phase fluctuations
coupled to the medium. 
At $(x,t)$, the probability amplitude for detecting 
a $K^0$ is
 
\begin{eqnarray}
\langle K^0\mid\psi(x,t)\rangle=\frac{1}{2}
\{{\rm exp}[{\rm i}(p_Lx-E_Lt)]+{\rm exp}[{\rm i}(p_Sx-E_St)]\}
\label{eq:eq6}
\end{eqnarray}
 
\noindent using $\langle K^0\mid K_L\rangle=\langle K^0\mid 
K_S\rangle=\sqrt{\frac{1}{2}}$. This is just as Eq. (\ref{eq:eq1}) 
above (with $\phi=0$), giving  the probability density as
 
\begin{eqnarray}
\mid\langle K^0\mid\psi(x,t)\rangle\mid^2=\frac{1}{2}\{
1+{\rm cos}\left[(p_L-p_S)x-(E_L-E_S)t\right]\}.
\label{eq:eq7}
\end{eqnarray}
 
Eq. (\ref{eq:eq7}) describes a plane-wave 
situation, with a unique value 
of $\sqrt{s}$, and with $x$ and $t$ as independent variables. 
In a realistic case, it would be used to form wave packets for
all particles, with a spread of values of $\sqrt{s}$. The size 
of such wave packets might be determined, for example, by the
time and position resolution of a detector in the incident beam 
or by the time structure of the accelerator beam. In any case, 
the packet cannot be larger than the $K_S$ lifetime, 
$\tau_S \sim 0.9 \times 10^{-10}s$, which gives
a spacial extent of typically about 2 cm. Thus the outgoing
kaon moves in a packet of this size, centered at the classical 
position. So the observation of the kaon at position $x$ must be 
made at a time when the packet is present, i.e. at time 
that is equal to $x/\beta$ or which differs from it by no more 
than the half-width of this wave packet. We therefore replace
$t$ in Eq. (\ref{eq:eq7}) with the time defined in this way using beta
calculated from the average $K_L$ and $K_S$ parameters:
  
\begin{eqnarray}
\beta=\frac{p_L+p_S}{E_L+E_S},
\label{eq:eq8}
\end{eqnarray}
 
\noindent which lies between the velocities $p_L/E_L$ and 
$p_S/E_S$. It is  crucial that the
observation is made at a {\it single} space-time point; no
meaning can be attached to the interference of wave functions
at different values of $x$ or $t$ (see \cite{hb,okun2}).
In any realistic case, the separation of the centres of the $K_L$ 
and $K_S$ wave packets is much smaller than their size, so in 
practice, there is no loss of coherence due to separation 
of the $K_L$ and $K_S$ packets.
 
Thus Eq. (\ref{eq:eq7}) becomes
 
\begin{eqnarray}
\mid\langle K^0\mid\psi(x,t)\rangle\mid^2=\frac{1}{2}\left[
1+{\rm cos}
\{(p_L-p_S)x-\frac{(E_L-E_S)}{\beta}x\}\right].
\label{eq:eq9}
\end{eqnarray}
 
\noindent Using
 
\begin{eqnarray}
\frac{1}{\beta}=
\frac{E_L^2-E_S^2}{(p_L+p_S)(E_L-E_S)}=
\frac{p_L-p_S}{E_L-E_S}+
\frac{2m\delta m}{(E_L-E_S)(p_L+p_S)},
\label{eq:eq10}
\end{eqnarray}
 
\noindent Eq. (\ref{eq:eq9}) becomes
 
\begin{eqnarray}
\mid\langle K^0\mid\psi(x,t)\rangle\mid^2=\frac{1}{2}\bigg[
1+{\rm cos}\frac{m~\delta m}{p}x\bigg]
\label{eq:eq11}
\end{eqnarray}
 
\noindent where $m$ and $p$ are the mean neutral kaon mass and
momentum. Thus the
wave number of strangeness oscillations in space is
$k=m~\delta m/p$. This is often expressed as an
oscillation in time, in which case, the angular frequency
is $\omega=m~\delta m/E$. Both of these results are in agreement
with the standard results \cite{bj}, without any assumption
about equality of momenta or energies.
 
\section{Discussion}
  
We conclude that a treatment of kaon and neutrino
oscillations, with the kinematics treated in full,
does indeed give the correct relation between
the mass difference and the oscillation wavelength. Most of 
the recent literature is in agreement with this standard result
for the wavelength,
though sometimes using incorrect kinematics. If CP violation and
the finite kaon lifetimes are incorporated in the above 
algebra, a more realistic result is obtained, with more
cumbersome-looking equations, but the wavelength of the
oscillations remains the same. The full equations are given
in \cite{jlowe,hb}.
 
Lipkin \cite{lip2} has suggested that the detector should also 
be included in the wave function of the system, since interaction
with it depends on the details of the neutrino or kaon wave
function. By choosing the kaon system here, we avoid this problem 
since the particles resulting from $K^0$ or $\bar{K}^0$ semileptonic 
decay identify the strangeness eigenstate, so no kaon detector, as 
such, is required. However, inclusion of a detector would not 
change our conclusions; at the zeros of the $K^0$ oscillation 
pattern, there are only $\bar{K}^0$ mesons and no $K^0$ mesons, 
so no detector could detect a $K^0$ at such a point.

If the mass eigenstates
did in fact have equal momenta or equal energies, then this
would imply a failure of 4-momentum conservation in the
kaon or neutrino production process. Such a failure would 
be evident well outside the oscillation region; the $K_L$,
which is the only state left in the asymptotic region, would,
in principle, have an energy or momentum inconsistent with 
4-momentum conservation.

In a recent preprint by Field \cite{field}, the kinematics and 
other aspects of the neutrino production are treated quite
differently, giving a wide range of correction factors to the
standard result for the wavelength. However, there is an error in
his derivation of the pion decay rates (Eq. (7) of Ref. \cite{field}).
When this is corrected, the motivation for his later modifications
to the standard treatment, 
with their bizarre physical consequences, is removed.
  
\ack{
This research is supported in part by the Department of Energy under
contracts W-7405-ENG-36 and DE-FG03-94ER40821, and by the NSF under
contract PHY-0099385.}

\end{document}